\RequirePackage{ifpdf}

\documentclass[prd,aps,
twocolumn,
nofootinbib,
floatfix,
superscriptaddress]{revtex4}
\usepackage{graphicx}
\usepackage{epsfig}
\usepackage{rotating}
\usepackage{amssymb}
\usepackage{dsfont}
\usepackage{psfrag}
\usepackage{latexsym,amsmath,
euscript,array,mathrsfs}

\def\k{\kappa}                    

\def\r{\rho}                                     

\def\6{\partial}

\newcommand{\be}{\begin{equation}}
\newcommand{\ee}{\end{equation}}
\newcommand{\beq}{\begin{equation}}
\newcommand{\eeq}{\end{equation}}
\newcommand{\bea}{\begin{eqnarray}}
\newcommand{\eea}{\end{eqnarray}}

\newcommand{\ba}{\begin{eqnarray}}
\newcommand{\ea}{\end{eqnarray}}

\newcommand{\beqs}{\begin{eqnarray}}
\newcommand{\eeqs}{\end{eqnarray}}
\newcommand{\bal}{\begin{aligned}}
\newcommand{\eal}{\end{aligned}}

\def\lbldef#1#2{\expandafter\gdef\csname #1\endcsname {#2}}

\def\href#1#2{#2}

\newcommand{\ber}{\begin{eqnarray}}
\newcommand{\eer}{\end{eqnarray}}

\newcommand{\beqar}{\begin{eqnarray}}

\newcommand{\eeqar}{\end{eqnarray}}

\newcommand{\dsl}
   {\kern.06em\hbox{\raise.15ex\hbox{$/$}\kern-.56em\hbox{$\partial$}}}

\newcommand{\eeqarr}{\end{eqnarray}}
\newcommand{\ZZ}{{\rm \kern 0.275em Z \kern -0.92em Z}\;}

\def\CC{{\mathchoice
{\rm C\mkern-8mu\vrule height1.45ex depth-.05ex
width.05em\mkern9mu\kern-.05em}
{\rm C\mkern-8mu\vrule height1.45ex depth-.05ex
width.05em\mkern9mu\kern-.05em}
{\rm C\mkern-8mu\vrule height1ex depth-.07ex
width.035em\mkern9mu\kern-.035em}
{\rm C\mkern-8mu\vrule height.65ex depth-.1ex
width.025em\mkern8mu\kern-.025em}}}

\def\RR{{\rm I\kern-1.6pt {\rm R}}}

\def\ZZ{{\rm Z}\kern-3.8pt {\rm Z} \kern2pt}
\def\IB{\relax{\rm I\kern-.18em B}}
\def\ID{\relax{\rm I\kern-.18em D}}
\def\II{\relax{\rm I\kern-.18em I}}
\def\IP{\relax{\rm I\kern-.18em P}}

\newcommand{\bear}{\begin{eqnarray}}
\newcommand{\eear}{\end{eqnarray}}

\def\k{\kappa}                    

\def\r{\rho}                                     

\def\6{\partial}

\begin{document}

\phantom{mi mono Amelio y yo}
\hfill MAD-TH-11-11

\title{Towards the string dual of  tumbling and cascading gauge theories}

\author{Eduardo Conde}
\affiliation{Departamento de  F\'\i sica de Part\'\i  culas\\ and\\ Instituto Galego de F\'\i sica de Altas
Enerx\'\i as (IGFAE), \\ Universidade de Santiago de Compostela\\E-15782, Santiago de Compostela, Spain}
\author{J\'er\^ome Gaillard}
\affiliation{Physics Department, University of Wisconsin-Madison, 1150 University Avenue\\
Madison, WI 53706-1390. USA}
\author{Carlos N\'u\~nez}
\affiliation{Swansea University, School of Physical Sciences, \\ Singleton Park, Swansea, Wales, UK}
\author{Maurizio Piai}
\affiliation{Swansea University, School of Physical Sciences, \\ Singleton Park, Swansea, Wales, UK}
\author{Alfonso V. Ramallo}
\affiliation{Departamento de  F\'\i sica de Part\'\i  culas\\ and\\ Instituto Galego de F\'\i sica de Altas
Enerx\'\i as (IGFAE), \\ Universidade de Santiago de Compostela\\E-15782, Santiago de Compostela, Spain}

\date{\today}


\begin{abstract}
We build type IIB  backgrounds that can be interpreted as the dual description
of field theories in which the dynamics shows many non-trivial phenomena,
generalizing the baryonic branch of the Klebanov-Strassler system.
We illustrate the steps  of the explicit construction with a particularly interesting example.
The dual field theory exhibits the expected behavior of an ${\cal N}=1$ supersymmetric gauge theory
which, over different ranges of the radial direction, is undergoing a cascade of Seiberg dualities,
a period of running, a cascade of Higgsings (tumbling) and finally confines with the formation of a  condensate.
 
\end{abstract}


\maketitle


%
%
%
%

\section{Introduction.}

The dynamics of four-dimensional field theories is  very rich, particularly at strong coupling.
Quantum effects give rise to very non-trivial renormalization-group flows
({\it running} of the couplings). At strong coupling many  theories
{\it confine}, and in some conditions gauge theories can undergo 
spontaneous symmetry breaking ({\it Higgsing}), possibly in multi-scale 
sequences ({\it tumbling})~\cite{tumbling}.
Under special circumstances (close to  approximate fixed points)
the running may be anomalously slow ({\it walking})~\cite{WTC}.
Finally, there are examples in which two different 
gauge theories (with different microscopic Lagrangians)
admit the same low-energy description
({\it Seiberg duality}) \cite{Seiberg:1994pq}. 
This feature can be iterated, giving rise to
what is called the {\it duality cascade}~
\cite{KS,Strassler:2005qs,Dymarsky:2005xt}.

On very general grounds, it is interesting  to have a (weakly coupled) dual gravity description of
(strongly coupled) field theories. The 
methods of the AdS/CFT correspondence~\cite{hep-th/9711200} 
generalize to cases where the dual field theory is not conformal~\cite{hep-th/9802042} and
 exhibits one or more of the dynamical features described above.
This allows to test quantitatively the properties of models for which the 
intuition based on perturbation theory fails to provide useful guidance.
Many such non-trivial features are believed to play important roles in
phenomenologically relevant models (for example, in dynamical electroweak-symmetry breaking~\cite{TC,ETC,reviews}), and the gravity duals offer
an opportunity to make quantitative predictions for measurable quantities.

The literature on the subject is very rich, see for 
instance~\cite{hep-th/9803131,KS,MN,BGMPZ}. 
In this Letter we take a further non-trivial step.
We present an algorithmic procedure that allows 
to construct a  large class of 
new backgrounds, dual to  ${\cal N}=1$ supersymmetric gauge theories.
Depending on some of the integration constants and parameters of the 
string configuration, the four-dimensional gauge theories
exhibit one or more of the features we referred to as running,  Higgsing, tumbling,
confinement and duality cascade.

We produce and discuss in some detail one special example 
of such a construction, by providing
 the essential technical  steps,  highlighting 
its preeminent physical properties and by 
commenting on generalizations of the construction itself.
For full details, we 
refer the reader to the vast literature on the subject and 
to a more extensive companion paper \cite{Conde:2011aa}.

\section{The backgrounds}

We consider  the class of type IIB backgrounds that can be 
obtained by the KK reduction of the theory on the base of a conifold, followed
by a consistent truncation. Our starting configuration contains
 $N_c$ units of flux
for the RR form $F_3$ (we call this the wrapped-D5 system),  to which we add
(fully back-reacted) $N_f$ smeared D5-brane sources.
There exists a procedure that, starting from this 
comparatively well-understood system, allows to generate 
a large class of backgrounds that are much more general, in which 
also the $H_3$ and $F_5$ forms are highly non-trivial and 
the dual field theory has a particularly rich multi-scale dynamics. 
Let us illustrate in this section how this is achieved.  

\subsection{The construction.}

Consider the ansatz for a background of type IIB string theory,
in which the ten-dimensional space-time 
consists of four Minkowski directions $x^{\mu}$, 
a radial direction $0\leq\r<+\infty$ and five internal angles $\theta,\tilde{\theta},\varphi,\tilde{\varphi},\psi$
parameterizing a compact manifold $\Sigma_5$~\cite{PT,Chamseddine:1997nm,Casero:2006pt,Casero:2007jj,HoyosBadajoz:2008fw}.
The full background is determined by a set of functions that are assumed to depend only on $\r$,
and by a set of non-trivial BPS equations that can be derived from  type IIB supergravity~\cite{Chamseddine:1997nm,MN,Casero:2006pt,Casero:2007jj}.

The whole dynamics controlling the background can be summarized by the following two equations~\cite{HoyosBadajoz:2008fw,Conde:2011rg,Barranco:2011vt}:
\begin{widetext}
\beqs
0&=& P'' + N_f\,S' + (\,P' + N_f\,S\,) \left( 
\frac{P' - Q' + 2 N_f\,S}{P + Q} + 
\frac{P' + Q' + 2 N_f\,S}{P - Q} - 4 \coth (2\r) \right) = 0\,\,,
\label{Eq:master}\\
Q & = & \coth(2\r)\,\,\Big[\,\int_0^\r dx \,
\frac{2N_c\,-\,N_f\,S(x)}{\coth^2(2x)}\,\Big]\,\,,
\label{Eq:Q}
\eeqs
\end{widetext}
where the primed quantities refer to derivatives with respect to $\r$. The  system is hence controlled
by the functions $P(\r)$, $Q(\r)$ and $S(\r)$. Once 
these functions are known, one can reconstruct 
 in a purely algebraic way
the whole
type IIB background, which consists of a non-trivial metric, 
dilaton and RR form $F_3$.

While the equations for $P$ and $Q$ are a 
repackaging of the BPS equations for the 
original ten-dimensional system, the function $S(\r)$ 
has a very different meaning: 
it controls the profile of the $N_f$ (flavor)
branes in the radial direction. For example, a vanishing $S$ 
yields the original (unflavored) wrapped-D5
system,  for which $\hat{P}=2N_c\r$ is a special 
solution~\cite{MN,Chamseddine:1997nm}.
On the other extreme, $S=1$ corresponds to the 
flavored solutions extensively discussed in the 
literature
~\cite{Casero:2006pt,Casero:2007jj,HoyosBadajoz:2008fw,arXiv:1002.1088}.

Following~\cite{Conde:2011rg,Barranco:2011vt}, 
we will assume that $S$ has a non-trivial $\r$-dependent 
profile, bounded by $0\leq S\leq 1$.
In particular, we will require that $S$ vanishes both in the deep IR (small $\r$)
and in the far UV (large $\r$), hence ensuring that asymptotically in the UV and in the IR the system 
resembles very closely the original wrapped-D5 system. The latter is the main element of
novelty of the proposal in this Letter. There are important, though subtle, differences
in the asymptotic expansions of backgrounds obtained for such a choice of $S$ 
with respect to the case $S=0$, as we will see.

In the case of the wrapped-D5 system (with $S=0$), $Q$ is exactly integrable, 
and the generic solution of Eq.~(\ref{Eq:master}) 
depends on two integration constants $c_{\pm}$,
which can be read from the UV expansion of the 
equation, see~\cite{HoyosBadajoz:2008fw}.
Their meaning in terms of the 
field-theory duals is well understood: they correspond to the insertion of
a dimension-eight operator (for $c_+$) and to a VEV for 
a dimension-six operator (for $c_-$), see \cite{Elander:2011mh}.

The choice of $c_{\pm}$ is not completely free. 
In particular, there exists a minimal value
for $c_+$, close to which the background solutions 
approach the special case $\hat{P}= 2 N_c \r$,
which does not admit a simple interpretation in terms of a local four-dimensional  
dual field theory.
We will hence avoid this case and $c_+$ will be kept explicit.
As for the second integration constant $c_-$, the fact that it is non-vanishing is connected with the 
appearance in the field theory of properties that resemble those of a walking field theory~\cite{NPP}.
Also, it seems that its presence ultimately produces 
a very mild singularity
in the deep IR. In this Letter, we will always keep $c_-=0$, so that the IR is
non-singular.

It is known~\cite{MM,GMNP,Elander:2011mh,rotations} that for backgrounds for which $c_+$ is non-trivial, there exists an
algebraic procedure (which we refer to as {\it rotation})
that allows to construct a new one-parameter family of solutions
starting from the wrapped-D5 system. In these new `rotated' 
backgrounds the warp factor $\hat{h}$ is given by
\beq
\hat{h}=1-\kappa^2 e^{2\Phi}=1-\k^2\sqrt{\frac{2e^{4\Phi_0}\sinh(2\r)^2}{(P^2-Q^2)(P'+N_f S)}}\geq0\,,
\eeq
where $\Phi$ is the dilaton\footnote{Notice that a factor of $4$ has been here 
reabsorbed in $e^{4\Phi_0}$ with respect to the notation 
in~\cite{GMNP,Elander:2011mh}, where also $\kappa$ is called $k_2$.}, and $\Phi_0$ and $\kappa$ are constants. We focus on cases where 
$P \sim c_+ e^{\frac{4}{3}\r}+{\cal O}(e^{-\frac{4}{3}\r})$ 
for $\r>\bar{\r}$, where the scale $\bar{\r}$ is determined by $c_+$ itself in a non-trivial way.
The dilaton  approaches a  constant $\Phi(\infty)$ at large $\r$
 (as opposed to being linear~\cite{MN}), hence
one finds the restriction $0\leq \kappa \leq e^{-\Phi(\infty)}$.

Summarizing, if we choose a non-zero\footnote{For $\kappa=0$ one recovers the original unrotated solution.} $\kappa$ and 
perform the rotation~\cite{MM,GMNP,Elander:2011mh,rotations}, this
yields a new type IIB background in which:
\begin{itemize}
\item the dilaton is unchanged,
\item the RR form $F_3$ is unchanged,
\item the Einstein-frame metric takes the form
\beqs
ds^2&=& e^{\Phi/2}\Big[ \hat{h}^{-1/2}dx_{1,3}^2 + \hat{h}^{1/2}ds_6^2\Big],
\eeqs
where $ds_6^2$ stands for the metric of the cone over the internal manifold
$\Sigma_5$,
\item the NSNS $H_3$  is non-trivial,
\item the self-dual RR field $F_5$ is non-trivial.
\end{itemize}

Before the rotation the background is controlled by $N_c$ D5-branes that 
are encoded in the flux of $F_3$, and by $N_f$ D5-branes that act as sources
and have a profile in the radial direction described by $S(\r)$.
After the rotation, besides these objects there is also a 
flux for $H_3$ which, together with the five-form flux $F_5$, encodes
background D3-branes and also a number of D3 sources 
induced on the $N_f$ 
D5 sources by the presence of the NSNS $B_2$ field.

The final step of the construction consists of choosing $\kappa$.
It has been observed in~\cite{Elander:2011mh} that the effect of the rotation is very dramatic in the 
far UV, but only sub-leading below the scale $\bar{\r}$ (in particular, $H_3$ and $F_5$ 
become very small below $\bar{\r}$).
The most interesting effect of the rotation is that it modifies the  warp factor
in front of the Minkowski part of the metric. In particular, by dialing $\kappa=e^{-\Phi({\infty})}$
one recovers a warp factor that approximates the one of the Klebanov-Strassler background,
i.e. a background that differs from being asymptotically AdS only by a logarithmic term.

In summary, we outlined a procedure that, given a function  $S(\r)$ 
that vanishes in the IR and in the UV (fast enough with $\r$), 
allows to construct a background which is 
free of any singular behavior. This background
has a metric very similar to the one of  (the baryonic branch of) the
Klebanov-Strassler solution \cite{KS},
but differs from it by the presence of an intermediate range 
of $\r$ where the  background
exhibits the presence of a distribution of D5 and D3 sources.
We will provide now a concrete example, before discussing the field-theory interpretation.

\subsection{A class of solutions.}

\begin{figure}[h]
\begin{center}
\begin{picture}(240,140)
\put(15,3){\includegraphics[height=4.5cm]{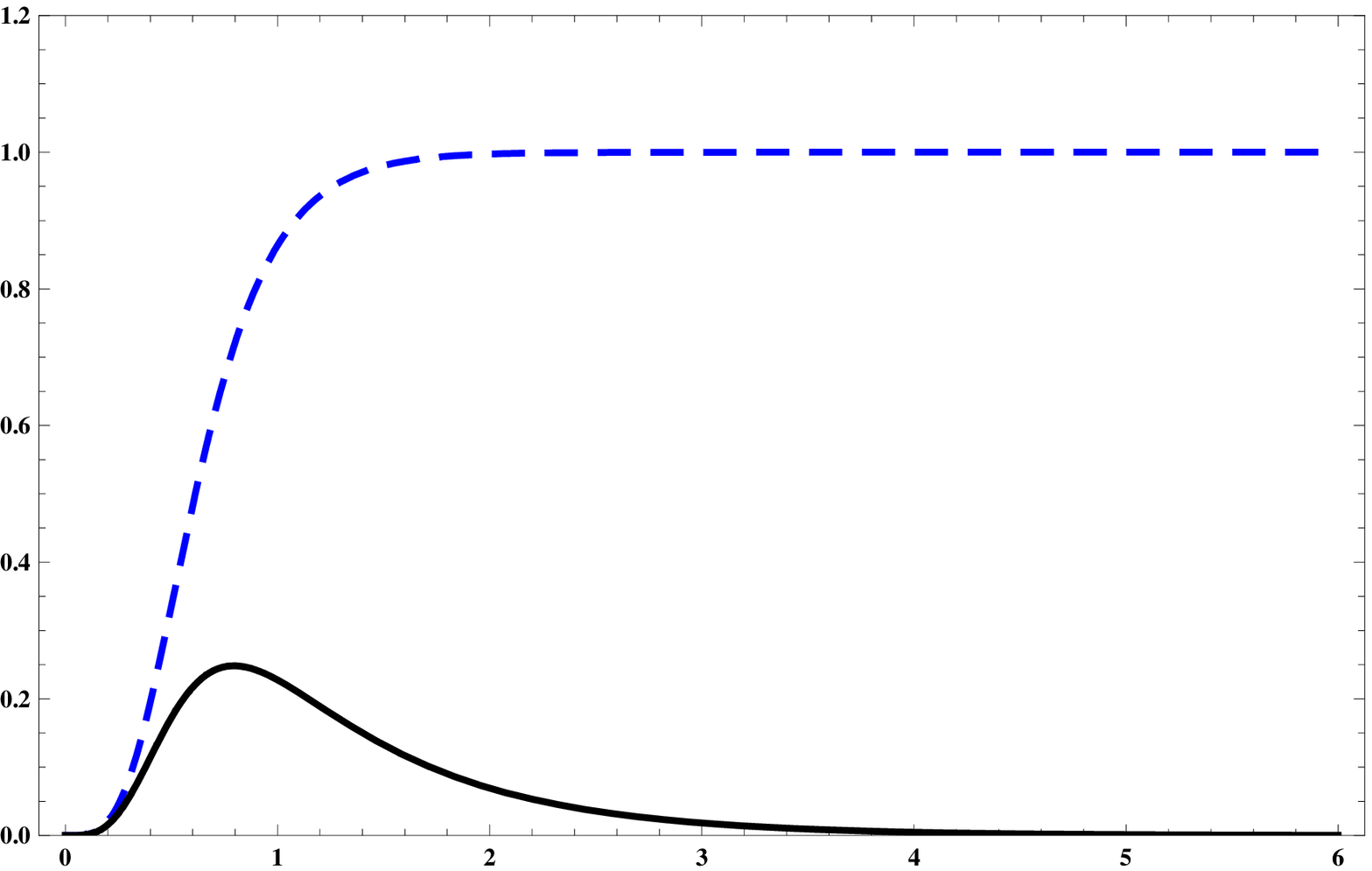}}
\put(190,0){$\r$}
\put(4,110){$S$}
\end{picture} 
\caption{The function $S$ used in this Letter (black), compared to a particular  one 
in~\cite{Barranco:2011vt} (dashed blue), which has $S=\tanh^4(2\r)$. }
\label{Fig:S}
\end{center}
\end{figure}

\begin{figure}[hb]
\begin{center}
\begin{picture}(240,140)
\put(15,3){\includegraphics[height=4.5cm]{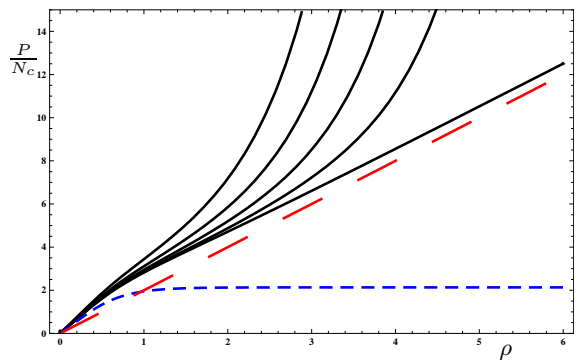}}
\put(190,0){$\r$}
\put(4,110){$\frac{P}{N_c}$}
\end{picture} 
\caption{The function $P$ solving the system defined here (black), 
compared to one
that solves the system for $S$ as in~\cite{Barranco:2011vt} (dashed blue) and to the original $\hat{P}$ 
solution, obtained for $S=0$ (long-dashed red). }
\label{Fig:P}
\end{center}
\end{figure}

We choose the following functional dependence for $S$
\beqs
S(\r)&= &\tanh^4(2\r) e^{-4\r/3}\,,
\label{Eq:S}
\eeqs
which we plot in Fig.~\ref{Fig:S}.
Before discussing the solutions, let us do some parameter counting.
We keep $N_c$ as a parameter, but set $\alpha^{\prime}g_s=1$.
The end of space is fixed at $\r_0=0$, which means that the function $S$ is positive, bounded 
and vanishes for $\r\rightarrow+\infty$ and 
for $\r\rightarrow 0$. We set $\Phi_0$ in such a way that 
$\Phi\rightarrow 1$ for $\r\rightarrow 0$ in all the numerical solutions we study.

Finally, the whole solution depends entirely on the value of $c_+$, since we choose 
the solution for $P$ to be linear in the IR (i.e. $P\sim h_1\r$), which 
necessarily implies $c_-=0$~\cite{HoyosBadajoz:2008fw,GMNP,Elander:2011mh}.
In the newly generated (rotated) background, 
the only new parameter is $\kappa$, which we fix to its 
maximal value $\kappa=e^{-\Phi(\infty)}$. 
We also set $N_f=2N_c$, and compare in the plots, where useful,  
to $S=\tanh^4(2\r)$~\cite{Barranco:2011vt}
 (in the following, we refer to the latter 
as the {\it monotonic} $S$).

We solve numerically for $P$, imposing that it be linear in the IR, and plot the result for
various possible solutions in Fig.~\ref{Fig:P}. 
 We find it convenient to compare
to the $\hat{P}=2 N_c \r$ solution (with $S=0$) 
and to a special solution $\bar{P}$ obtained with the monotonic $S$.
The class of solutions we find consists 
of the black curves in Fig.~\ref{Fig:P}. 
The comparison shows that deep in the IR our $P$ 
agrees with $\bar{P}$. However, at large $\r$ one recovers a behavior 
similar to $\hat{P}$, at least until the scale $\bar{\r}$ at which the exponential behavior appears.

Notice that this scale $\bar{\r}$ can be 
indefinitely increased, hence one might 
envision the case in which the asymptotics resembles that of $\hat{P}$. For this limiting case, we can approximate in the UV
\beq
P\simeq2N_c\rho+1-2S_{\infty}\,,
\label{Eq:asymp}
\eeq
neglecting terms that vanish at large $\r$, where the constant $S_{\infty}$ is given by
\beqs
S_{\infty}&=&\int_0^{+\infty} d \r \tanh^2(2\r)S(\rho)\,\simeq\,0.29\,.
\eeqs
We see that the non-trivial profile of $S$ shifts $P$ by a constant amount 
with respect to $\hat{P}=2N_c \r$ in the far UV.

More interesting for our purposes are the cases in which the UV asymptotic behavior of $P$ is
exponential, because in this case we can perform the rotation procedure.
As a result, the asymptotic behavior of the solution is better 
understood by looking at the
warp factor $\hat{h}$, which for large $\r$ is 
\beq
\hat{h}=\frac{3N_c^2e^{-\frac{8}{3}\r}}{8c_+^2}\left(8\r-1+\frac{2c_+N_f}{N_c^2}-\frac{4N_f}{N_c}S_{\infty}\right)+{\cal O}(e^{-4\r})\,.
\eeq
This expression shows two very interesting facts: first of all, as anticipated, one obtains the familiar
result that the warp factor is almost AdS, except for the term linear in $\r$ in parenthesis.
Notice we reinstated $N_f$ in the expansions for clarity, but one should keep in mind that in the plots we explore the case $N_f=2N_c$.

Concluding, let us summarize what are the properties of the generic $P$ belonging to this one-parameter family. Deep in the IR, it looks 
somewhat similar to $\hat{P}=2N_c \r$~\cite{MN}, 
but with a slightly larger slope\footnote{
This different slope is just due to the specific functional 
dependence of $S$:  one recovers the same slope as in $\hat{P}=2N_c\r$ 
by replacing 
the $\tanh^4(2\r)$ factor  by $\tanh^{2N}(2\r)$ 
with $N$ very large~\cite{Barranco:2011vt}.}.
Over the range in which $S$ is non-trivial, 
$P$ keeps growing, but with a smaller derivative.
This  range gives way at intermediate values of $\r$ 
to a $P$ which has a slope very close to $\hat{P}=2N_c\r$,
but which is shifted upwards
 by a finite amount, as reflected in Eq.~\eqref{Eq:asymp}. Finally, in the far UV, $P$ gives rise (after the rotation is performed) to a warp
factor that is very close to the one of the KS case, but again with a shift in
the logarithmic term.
We will now go on to suggest an interpretation for 
these findings in field-theory terms.

\begin{figure}[h]
\begin{center}
\begin{picture}(220,280)
\put(17,3){\includegraphics[height=4.5cm]{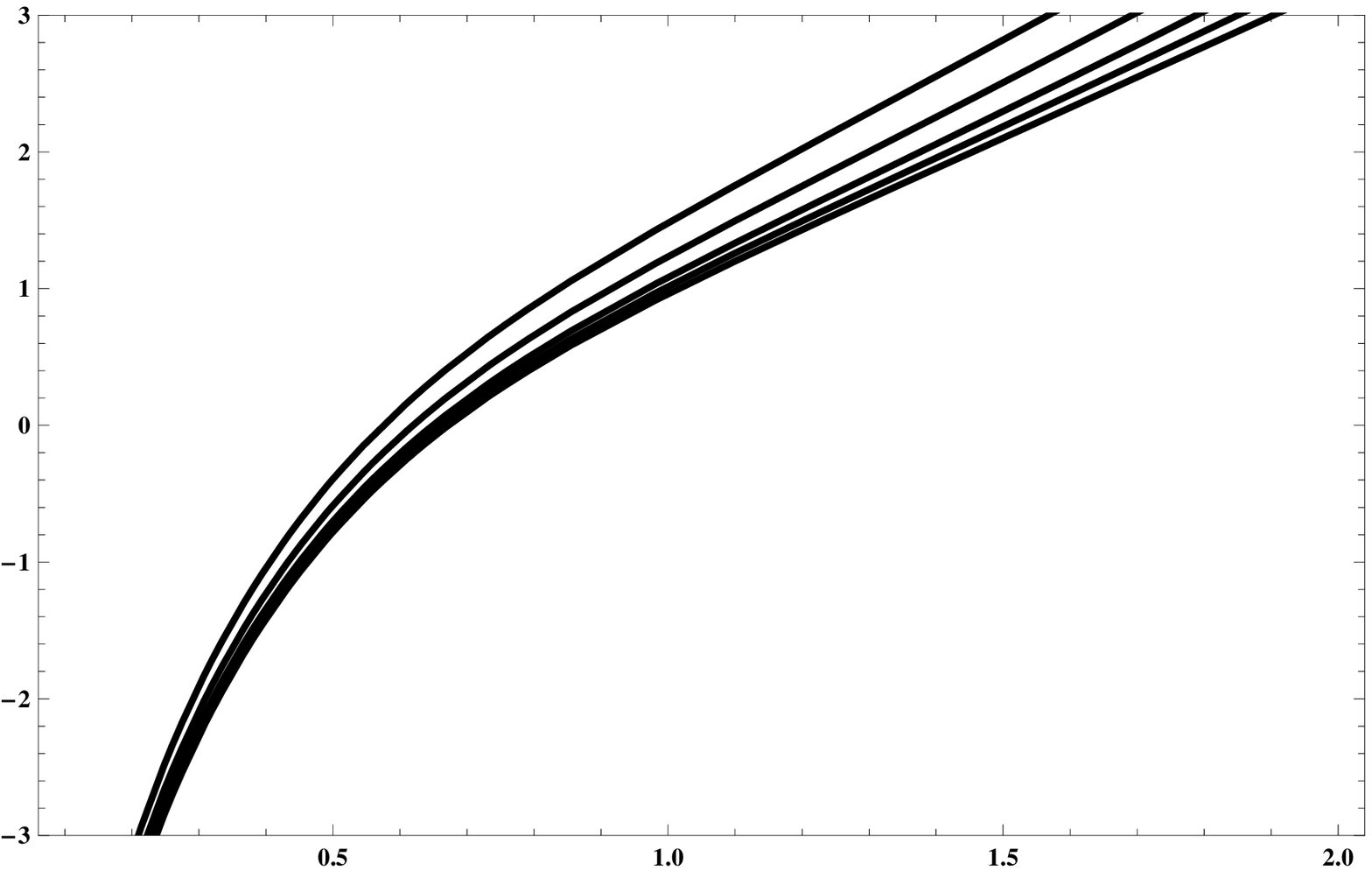}}
\put(15,143){\includegraphics[height=4.5cm]{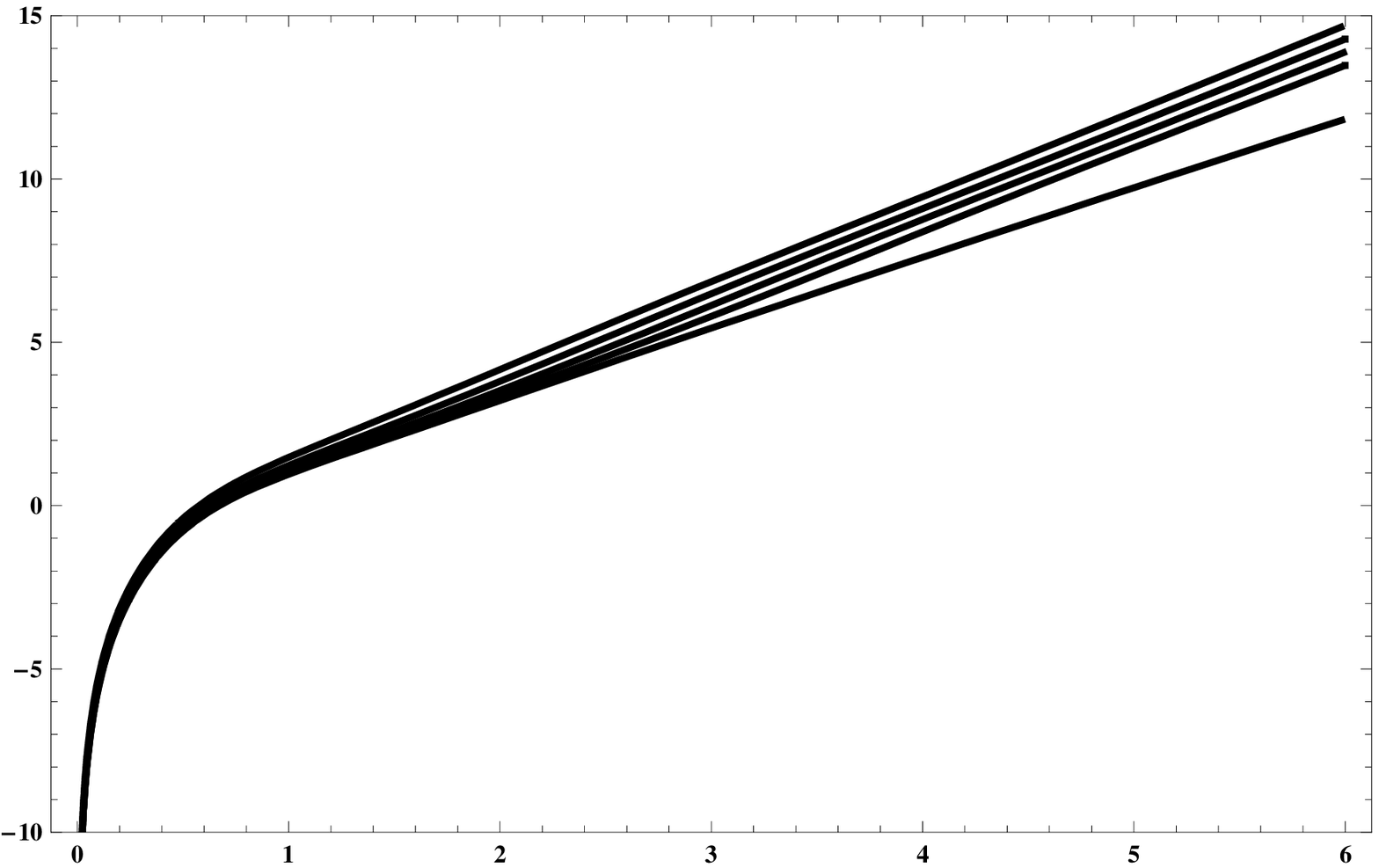}}
\put(190,0){$\r$}
\put(1,110){$\log c$}
\put(190,140){$\r$}
\put(1,252){$\log c$}
\end{picture} 
\caption{The central charge $c$ obtained with the functions $P$ solving the system defined in the Letter,
and shown as black curves in Fig.~\ref{Fig:P}. }
\label{Fig:c}
\end{center}
\end{figure}

\section{Field-Theory interpretation.}

The first thing we want to study in order to provide a sensible field-theory interpretation is
the central charge $c(\r)$ of the rotated system, which turns out to be 
\beqs
c &=& \frac{\hat{h}^2\, e^{2\Phi} (P^2 -Q^2) (P'+N_f S)^2}{128 [\partial_{\rho} \ln (\sqrt{\hat{h}}\, e^{2\Phi} (P^2 -Q^2) \sqrt{P'+N_f S})]^3}\,.
\label{Eq:c}\\ \nonumber
\eeqs
The results are illustrated in Fig.~\ref{Fig:c}. The central charge is 
positive-definite and
monotonically increasing towards the UV,
as expected from consistency with the $c$-theorem.
It vanishes near the end of space, as
expected because the dynamics ultimately yields confinement. 
Its behavior is regular, with sharp 
changes at the scale $\bar{\r}$ and in the region where $S\neq 0$ (see in particular the detail of this region
in Fig.~\ref{Fig:c}).
Notice that while qualitatively similar, the plots obtained for various 
choices of $P(\r)$ are different, and agree only when $\r\rightarrow 0$.

The second important quantity we want to use is the Maxwell charge of the theory, 
which (after the rotation) counts the 
total D3 charge in the background; this is made out of (flux) D3 charge 
plus the source D3 charge
induced  as an effect of the NSNS $B_2$ field on the world-volume of the 
D5 sources.

We define the number $n$ of bulk D3-branes and $n_f$ of 
source D3-branes as given by
\begin{widetext}	
\beqs
		n + n_f &\equiv &\int_{\Sigma_5} F_5 = \int_{\Sigma_5} B_2 \wedge F_3\,=\nonumber\\
		&=& 4 \pi^3 e^{\Phi(\infty)} N_c \left[ \frac{N_f}{N_c} P S + 2 Q \left(1 - \frac{N_f}{2N_c} S\right) \tanh(2\rho) - \frac{4}{N_c} \frac{Q^2}{\sinh(4\rho)} \right]\,.
		\eeqs
\end{widetext}

\begin{figure}[t]
\begin{center}
\begin{picture}(220,280)
\put(27,143){\includegraphics[height=4.5cm]{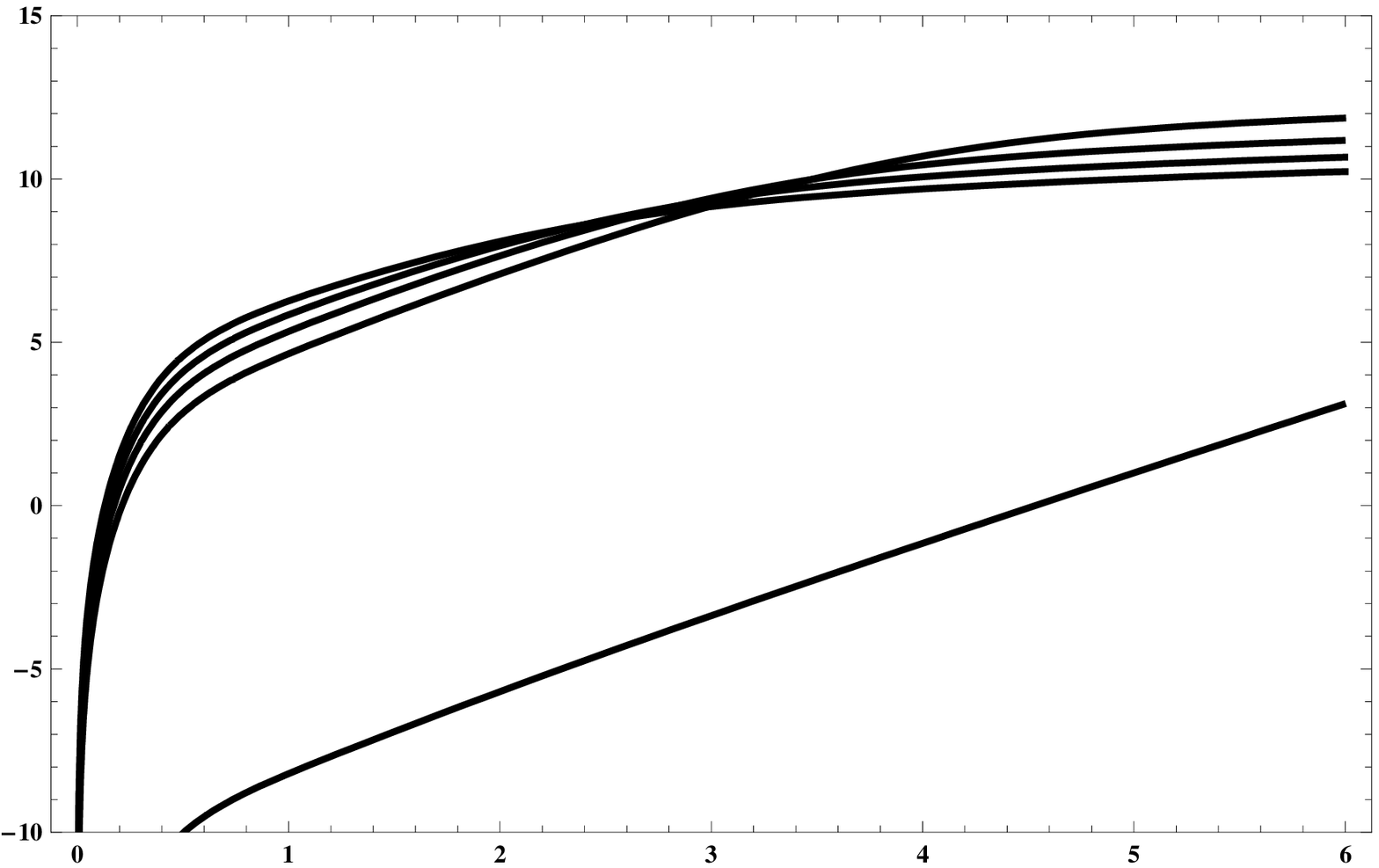}}
\put(23,3){\includegraphics[height=4.5cm]{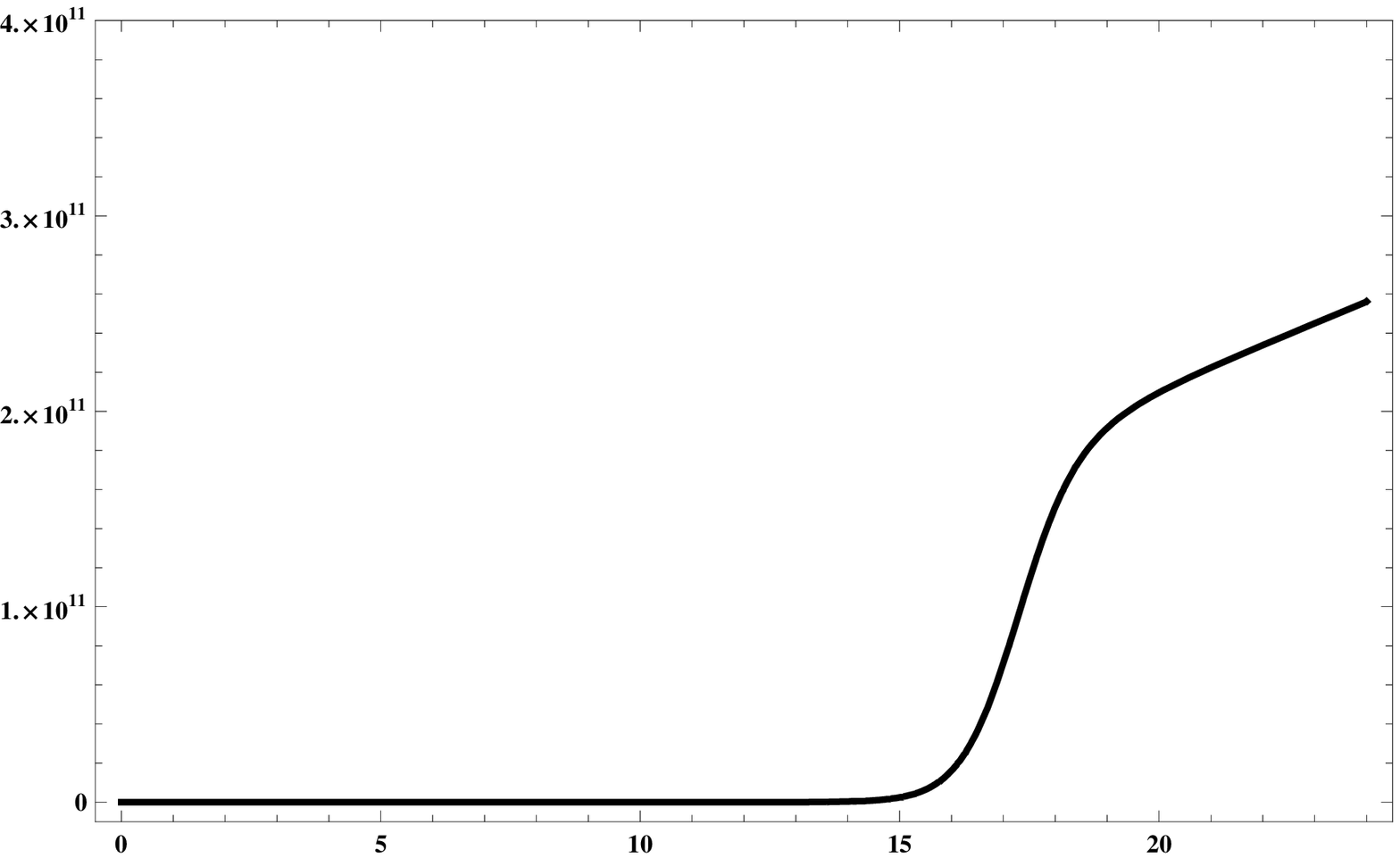}}
\put(205,0){$\r$}
\put(-12,110){$ (n+n_f)$}
\put(205,140){$\r$}
\put(-12,250){$\log (n+n_f)$}
\end{picture} 
\caption{The function $\log(n+n_f)$ obtained with the functions $P$ solving the system defined in the Letter,
and shown as black curves in Fig.~\ref{Fig:P}. Notice that the lower curve in this plot corresponds to the one that looked linear in that figure. The lower plot shows $n+n_f$ for the numerical
solution with largest $\bar{\r}$, which clearly exhibits the sudden drop at scale $\bar{\r}$, together with the linear dependence when $\r>\bar{\r}$.}
\label{Fig:npnf}
\end{center}
\end{figure}

The result of this is plotted in Fig.~\ref{Fig:npnf},  for the same numerical solutions as in Fig.~\ref{Fig:P} and
in Fig.~\ref{Fig:c}.
The resulting functions are positive-definite, monotonically increasing, and vanish at the end of space,
as required by consistency.
The most visible property of $n+n_f$ is that it drops towards zero below the scale $\bar{\r}$, 
which was already noticed in~\cite{Elander:2011mh}
 for $N_f=0$, and interpreted in terms of the Higgsing of the 
quiver into a single-site model. For $\r>\bar{\r}$, one sees that $n+n_f$ is linear in $\r$:
\beq
\frac{n+n_f}{4\pi^3N_c^2}\simeq e^{\Phi(\infty)}\left[4\r-2+\frac{N_f}{N_c}
\left(\frac{c_+}{N_c}-2S_{\infty}\right)\right]\,.
\eeq

We are now going to propose a possible interpretation for the dual backgrounds.
\begin{itemize}
\item At large scales ($\r>\bar{\r}$), the dual theory is a quiver with gauge group $SU(n+n_f+N_c)\times SU(n+n_f)$, where $n=k N_c$ is related to the D3 charge of the background, and $n_f$ to the D3
 sources induced by the field $B_2$ on 
the D5 sources. The theory is undergoing a cascade of Seiberg dualities 
 of the form 
 \beq\bal
& \qquad SU(n_f+(k+1)N_c)\times SU(n_f+kN_c)\\
&\rightarrow SU(n_f+(k-1)N_c)\times SU(n_f+kN_c)\\
&\rightarrow SU(n_f+(k-1)N_c)\times SU(n_f+(k-2)N_c)\,\rightarrow\, \cdots\,.
\eal\eeq

\item At the scale  $\bar{\r}$ the cascade stops abruptly,
and the quiver theory is Higgsed down to a single-site 
theory because of the formation of a dimension-two condensate, which is related to the  operator $\cal U$
defined in~\cite{Dymarsky:2005xt}.
As discussed in~\cite{AD,MM,Dymarsky:2005xt,Elander:2011mh}, the arbitrariness of $\bar{\r}$ is loosely
related to the modulus that controls the dimensionality of
the coset. The gauge group reduces to $SU(n_f+N_c)$.
For energies below the scale $\bar{\r}$, the original wrapped-D5 system yields a good 
effective-field-theory description.

\item In the intermediate range where $\r<\bar{\r}$, but where $S\simeq 0$, the theory is 
an ${\cal N}=1$ SUSY theory with gauge group $SU(n_f+N_c)$ and massless matter fields.
Besides these, as a result of the Higgsing, the spectrum of massive states deconstructs  
two extra dimensions, so that the theory apparently 
looks like a higher-dimensional 
field theory~\cite{AD}. 

\item In the energy range where $S$ is non-vanishing, 
a cascade of Higgsings is taking place. The idea
\cite{hep-th/0101013} is that
for every source brane that is crossed
when flowing down in the radial direction we Higgs the groups. 
This
sequentially reduces further the gauge group to $SU(N_c)$, 
while at the same time giving mass to the 
matter-field content. The smooth behavior of $S$ at large $\r$ 
yields small threshold effects, that
are the reason why the $\hat{P}$ solution is not exactly 
reproduced, as we commented earlier.

\item Very deep in the IR, close to the end of space, 
the  theory finally resembles
${\cal N}=1$ SYM  with $SU(N_c)$ group in four dimensions,
confines and produces a non-trivial 
gaugino condensate.

\end{itemize}

The dual field theory is showing, at different scales, the behavior of a 
field theory that undergoes a cascade of Seiberg dualities, the Higgsing of the quiver theory 
due to the $\cal U$ condensate, a tumbling sequence which reduces the rank of the one-site 
field theory, and finally confinement and  
gaugino condensation.

\section{Outlook}

There are a number of possible future research directions that can be developed
starting from the results we summarized here and in a companion paper
\cite{Conde:2011aa}.

In this Letter we presented one very special class of type IIB backgrounds that exhibit
some of the properties expected in field theories that have a very interesting and rich 
dynamics. 
Some of the statements we made about the interpretation in field-theory terms 
are mostly based on circumstantial evidence, and it would be useful to 
find other ways to check whether our interpretation is correct.
In particular, this means that one would like to explore in a more systematic
way the space of acceptable profiles for the function $S$, verify that the resulting
backgrounds are consistent, and test whether a generalization of the 
arguments we summarized here still provides a satisfactory explanation of the results.
In particular, it would be interesting to see what happens when the support of $S$ is very large,
and far from the end of space of the geometry.

On the more phenomenological side, this is an interesting step in the direction of studying tumbling
dynamics, which is believed to have an important role in the context of dynamical electroweak-symmetry breaking. It would be interesting to find models of this type that resemble as much as possible 
phenomenologically viable scenarios, and use them to compute quantities that 
are relevant to modern high-energy Physics.

\section*{Acknowledgments}

Discussions with various colleagues
helped to improve the content and presentation of this Letter.
We wish to thank Lilia Anguelova, Stefano Cremonesi, Anatoly Dymarsky, Daniel Elander, Aki Hashimoto, Tim Hollowood, Prem Kumar, Dario Martelli, Ioannis Papadimitriou and Jorge Russo.
The work of J.G. was funded by the DOE Grant DE-FG02-95ER40896. The  work of E.~C. and A.~V.~R.  was funded in part by MICINN   under grant
FPA2008-01838,  by the Spanish Consolider-Ingenio 2010 Programme CPAN (CSD2007-00042) and by Xunta de Galicia (Conseller\'\i a de Educaci\'on and grant INCITE09 206 121 PR) and by FEDER. E.~C. is supported by a Spanish FPU fellowship, and thanks the FRont Of Galician-speaking Scientists for unconditional support.

\end{document}